\documentclass[aps,prd,preprintnumbers,showpacs,showkeys,nofootinbib,
superscriptaddress,fleqn,floatfix,tightenlines,twocolumn]{revtex4-1}
\usepackage{amsmath,amsfonts,amssymb,amscd,amsxtra,amsthm}
\usepackage{graphicx}  
\usepackage{epstopdf}
\usepackage{dcolumn}  
\usepackage{bm}          
\usepackage{slashed}
\usepackage[utf8]{inputenc} 

\newcommand{\intrT}{\pi\int\limits_0^\infty\!dr\,r^{2}
\!\int\limits_0^\pi\!s_\theta\,d\theta\,}

\usepackage[normalem]{ulem} 
\usepackage[dvipsnames]{xcolor} 
\renewcommand\sout{\bgroup \color{red} \ULdepth=-.5ex \ULset}

\newcommand\T{\rule{0pt}{2.6ex}}       

\begin{document}
\preprint{INHA-NTG-01/2019}
\title{Nucleon and $\Delta$ isobar in a strong magnetic field} 

\author{Ulugbek~Yakhshiev}
\email{yakhshiev@inha.ac.kr}
\affiliation{Department of Physics, Inha University,
Incheon 22212, Republic of Korea}

\author{Hyun-Chul~Kim}
\email{hchkim@inha.ac.kr}
\affiliation{Department of Physics, Inha University,
Incheon 22212, Republic of Korea}
\affiliation{Advanced Science Research Center, Japan Atomic Energy
  Agency, Shirakata, Tokai, Ibaraki, 319-1195, Japan} 
\affiliation{School of Physics, Korea Institute for Advanced Study
  (KIAS), Seoul 02455, Republic of Korea} 

\author{Makoto Oka}
\email{oka@post.j-parc.jp}
\affiliation{Advanced Science Research Center, Japan Atomic Energy
  Agency, Shirakata, Tokai, Ibaraki, 319-1195, Japan} 

\date{\today}

\begin{abstract}
We investigate the static properties of the nucleon in the presence of
strong magnetic fields and discuss the consequent changes of the
nucleon structure, based on the Skyrme model.  
The results show that at large values of the 
magnetic field ($\sim10^{17}$ to $10^{18} \mathrm{G}$), which is
supposed to appear in heavy-ion collision experiments at RHIC
energies, the soliton starts to deviate from the spherically symmetric
form and its size starts to change. At extremely large 
values of the magnetic field ($\sim 10^{19}$ G),   
which may be found at the LHC experiments,  
the soliton becomes more compact than in free space. 
The results also show that in the presence of the external 
magnetic field, the mass of the nucleon tends to increase in general
and the mass degeneracy of the $\Delta$ isobars from isospin symmetry 
will be lifted. We also discuss the changes in the 
mass difference between the $\Delta$ and the nucleon, $\Delta
m_{\Delta\mathrm{N}}$, due to the influence of the external magnetic
field. We  find that $\Delta m_{\Delta\mathrm{N}}$  increases as the
strength of the magnetic field grows.
\end{abstract}


\maketitle
\section{Introduction}

Understanding how hadrons are modified in the presence
of various external fields is an important topic in contemporary
physics of hadrons. In particular, it is of great interest
to investigate how the nucleon undergoes change in a 
strong magnetic field, since it provides certain information  
on both compact astrophysical objects and ultra-relativistic heavy-ion
collision (URHIC), which unveils the nature of matter in the early
Universe. A very strong magnetic field may exist in a
magnetar in which the magnetic field reaches an order of $B_M\sim   
(10^{11}-10^{15})$~G~\cite{Mereghetti:2015asa,
  Kaspi:2017fwg}\footnote{Here $B_M$ denotes the strength of a
  magnetic field. Note that we adopt the gauss ($\mathrm{G}$) as the
  unit of the strength of the magnetic field. 1 G corresponds to
  $2\times 10^{-20}\,\mathrm{GeV}^2$.}.  
Even stronger magnetic fields ($\sim 10^{16}$ to $10^{17}$ G) may be
found in the cosmological $\gamma$-ray
bursts~\cite{Kouveliotou:1998ze, Bhattacharjee:1998qc,Thompson:1996pe}.  
However, one
can create even much stronger magnetic fields in the course of
relativistic heavy-ion collisions~\cite{Rafelski:1975rf}.
At the Relativistic Heavy-Ion Collider (RHIC), the magnetic
field could reach $B_M\sim 3\times
10^{18}$\,G and it may even rise to $B_M\sim 10^{19}$\,G at
the Large Hadron Collider(LHC)~\cite{Skokov:2009qp, Huang:2015oca, 
  Kharzeev:2007jp,Voronyuk:2011jd, Ou:2011fm, Bloczynski:2013mca,
  Deng:2014uja, Hattori:2016emy, Zhao:2017rpf}.    
Although such an extremely strong magnetic field exists only during a
very short period of time, it may bring about the distortion of hadrons
and may change their properties greatly.

There has been already a great deal of theoretical works on
modifications of hadrons under the influence of strong magnetic 
fields~\cite{Hidaka:2012mz, Machado:2013rta, Alford:2013jva,Machado:2013yaa,Luschevskaya:2014lga,Luschevskaya:2015bea, Taya:2014nha, Bonati:2015dka, Gubler:2015qok,Suzuki:2016kcs, Andreichikov:2016ayj, Liu:2018zag,Coppola:2018vkw,Avancini:2018svs,Avancini:2016fgq, Andreichikov:2013pga}. 
  However, while they mainly concentrate
on the modification of light and heavy meson properties in the presence
of the strong magnetic fields, there are only few works on the changes
of properties of the nucleon~\cite{Andreichikov:2013pga, 
  Bigdeli:2017uzy}. Since Refs.~\cite{Andreichikov:2013pga, 
  Bigdeli:2017uzy} aim at describing the neutron stars, they  
focus only on the modification of the neutron in the strong magnetic
fields. In the present work, we will investigate the modifications of
the nucleon and $\Delta$ isobar properties in the presence of the
strong magnetic fields within the framework of a chiral soliton
approach.  

The approach provides a simple but effective way of
describing the structure of the nucleon. The main idea arises from the 
seminal papers by Witten~\cite{Witten:1979kh, Witten:1983tw,
  Witten:1983tx}. In the limit of $N_c\to \infty$ ($N_c$ as the
number of colors), the mass of the nucleon is proportional to $N_c$
whereas its width is of order $\mathcal{O}(1)$, which indicates that
the meson fluctuations can be neglected. In this picture, a baryon
arises as a topological chiral soliton that is called
skyrmion~\cite{Skyrme:1961vq, Adkins:1983ya}. 
The nucleon as a chiral soliton is naturally an extended object, so
that one can examine how the nucleon undergoes 
changes when a very strong magnetic field is exerted on it. A
theoretical method has been developed over years, the environment
surrounding the nucleon being treated collectively. It has been
successfully applied to the description of the nucleon 
in nuclear medium~\cite{Kim:2012ts, Jung:2012sy, Yakhshiev:2013goa, 
Jung:2014jja, Jung:2015piw, Hong:2018sqa},
the nucleon in finite nuclei~\cite{Yakhshiev:2001ht},
the properties of nuclear matter~\cite{Yakhshiev:2013eya}  
and even to the explanation of properties of atomic 
nuclei~\cite{Meissner:2008mr, Meissner:2009hm}. The similar
theoretical tool can be utilized for describing the nucleon in the
strong magnetic field.  

From a technical point of view, the nucleon in an external magnetic
field is very similar to the situation when a skyrmion is
embedded into an isospin asymmetric nuclear
environment~\cite{Yakhshiev:2013eya}. In general, 
one may expect that the magnetic field will change the
nucleon properties less than the effects of isospin symmetry breaking.
However, when it comes to the very strong magnetic fields that reach
the level of URHICs at the LHC, the effects from the magnetic fields
may become sizable.  In this case, they may also play a crucial role 
in describing the evolution of the universe at an early
stage~\cite{Steigman:2005uz,Cyburt:2004cq}. Moreover,   
such strong magnetic fields will reveal certain novel features
relevant to the structure of the nucleon.  

Depending on a specific configuration of the external magnetic field, 
one may further expect possible nonspherical
deformations of the skyrmion in isospin and ordinary 
spaces~\cite{Yakhshiev:2001ht, Meissner:2008mr, Meissner:2009hm}  
from the spherically symmetric \emph{hedgehog} form corresponding to
the skyrmion in free space~\cite{Adkins:1983ya,Adkins:1983hy}.
In this sense, the situation becomes even more interesting if the 
nucleon properties are studied in the presence of external 
isospin asymmetric nuclear environment that actually creates the 
strong magnetic field, that is, if the nucleon is located 
inside compact stellar objects in the presence of strong magnetic 
fields. The corresponding investigation can naturally be performed by
generalizing the approach developed in Refs.~\cite{Yakhshiev:2013eya,
  Yakhshiev:2001ht, Meissner:2008mr, Meissner:2009hm}  
in the presence of an additional external magnetic field. However, we
will concentrate only on  the external magnetic field for simplicity
and leave more general and complex studies as future works. 

In the present work, we consider the homogeneous magnetic field
oriented along the axis of quantization. This choice allows us to
consider axially symmetric solutions of the classical equation of
motion for the soliton instead of a complicated situation where the 
soliton has totally an asymmetric form. Then we can employ the
technique developed already for asymmetric nuclear 
environment~\cite{Yakhshiev:2001ht, Meissner:2008mr,
  Meissner:2009hm}. Nevertheless, it is necessary to note that in the
present work  there will be some differences at the Lagrangian level
due to the nature of the external magnetic field influencing the 
properties of the nucleon under consideration.
In Refs.~\cite{Yakhshiev:2001ht,Meissner:2008mr,Meissner:2009hm}
the effect of environment on the skyrmion properties 
was introduced by means of the density functions, based on
phenomenological information taken from mesonic atoms at low
densities. Further modifications were achieved by introducing 
another density functions into the Lagrangian 
and relating them to the properties of nuclear matter near the
saturation point
$\rho_0\approx0.16$\,fm$^{-3}$~\cite{Yakhshiev:2013eya}.  
In the present work, the external magnetic field will be introduced
by taking into account the U(1) gauge field into the
original effective chiral Lagrangian~\cite{Gasser:1984gg}. 
 
The present paper is organized as follows: In the 
next Section~\ref{sec:Lag&ansatz}, we briefly discuss the Lagrangian
of the model and the axially symmetric ansatz for the solutions of 
field equations.  In Section~\ref{sec:clmass}, we explain the
variational method for the problem and discuss the parametrizations
of profile functions. We also discuss the minimization process and 
present the classical results. Then we discuss how the 
baryon charge distribution is changed to a spheroidal form under the
influence of the magnetic field. In Section~\ref{sec:quant}, we show
how to quantize the spheroidal skyrmion and discuss the changes of the
nucleon properties in the magnetic field.  In the last
Section~\ref{sec:summary&outlook} we  summarize the present results,
draw conclusions, and give future outlook.
The explicit expressions of the mass functional and the
moments of inertia of the spheroidal skyrmion can be  
found in Appendix~\ref{app:mass_momin}.
\section{Lagrangian and ansatz}
\label{sec:Lag&ansatz}
We start with the effective chiral Lagrangian,  
incorporating explicit chiral symmetry breaking~\cite{Adkins:1983hy}
\begin{align}
{\mathcal L}=&-
\frac{F_{\pi}^2}{16}\mathrm{Tr}\,L_\mu L^\mu
               +\frac{1}{32e^2}\mathrm{Tr}[L_{\mu},L^{\nu}]^2\cr 
&+\frac{F_{\pi}^2m_{\pi}^2}{16}
\mathrm{Tr}[U+U^{\dagger}-2]\,,
\label{lag}
\end{align}
where the first term is called the Weinberg term and 
the second one was originally introduced by
Skyrme~\cite{Skyrme:1961vq}, which is also known as the
Gasser-Leutwyler term in the large $N_c$. The chiral current $L_\mu$
is defined as $L_\mu=U^+\partial_\mu U$, where the 
SU(2) unitary matrix $U=\exp\{2i\tau_a\pi_a/F_\pi\}$ is expressed in
terms of the Cartesian isospin-components of the pion field
$\pi_a~(a=1,2,3)$. $\tau^a$ stand for the Pauli matrices in isospin
space. There are three input parameters, i.e. the pion decay constant
$F_\pi =108.783$\,MeV, the Skyrme parameter $e=4.854$,  
and the pion mass $m_\pi = 134.977$\,MeV, which are chosen in such a
way that the model properly reproduces the experimental data on the
masses  of the proton and neutron with breakdown of isospin symmetry
taken into account (for the details,  see
Refs.~\cite{Meissner:2008mr,Meissner:2009hm}). 

In order to consider the effects of the external magnetic field we
introduce the U(1) gauge field into the Lagrangian of
Eq.\,(\ref{lag}).  So, the ordinary derivative is replaced by the
covariant one given in the form of   
\begin{align}
D_\mu U=\partial_\mu U + iq_eA_\mu[Q,U],
\end{align}
where $q_e$ denotes the electric charge and $A_\mu$ stands for the
electromagnetic four-vector potential (for example,
see~Ref.\cite{Rudy:1994qb}). 
Here the charge operator in the SU(2) framework is defined as 
\begin{align}
Q=\frac{1}{6}\,\mathbb{I}+\frac{1}{2}\,\tau_3\,.
\end{align}

As mentioned above, we introduce the homogeneous magnetic field along 
the quantization axis or the $z$ direction $\bm{B}_M=(0,0,B_M)$, so
we fix correspondingly the gauge of $A^\mu$ as follows 
\begin{align}
A^\mu=\left(0,-\frac{1}{2}y B_M,\frac{1}{2} x B_M,0\right).
\end{align}
When the magnetic field is absent, the hedgehog ansatz is imposed to
be a spherically symmetric hedgehog form $U=\exp\{i\bm{\tau}\cdot
\bm{n}P(r)\}$, where the unit vector in isospin space is chosen as a
normal vector $\bm{n}$ in ordinary three dimensional space. However,  
the ansatz for the skyrmion in the presence of the magnetic field may
be deformed in the isospin and ordinary spaces deviating from 
the original spherical form in the absence of external fields.  
The most general form of the ansatz, which takes into account
all possible deformations, can be represented as 
\begin{align}
U(\bm{r})= \exp\left\{i\bm{\tau}\cdot\bm{N}(\bm{r})
P(\bm{r})\right\}
\label{ansatz}
\end{align}
where the normal vector in isospin space is expressed as 
\begin{align}
\bm{N}=\left(
\begin{array}{c}
\sin\Theta(r,\theta,\varphi)\cos\Phi(r,\theta,\varphi)\\
\sin\Theta(r,\theta,\varphi)\sin\Phi(r,\theta,\varphi)\\
\cos\Theta(r,\theta,\varphi)
\end{array}\right)
\end{align} 
in terms of two profile functions, $\Theta(r,\theta,\varphi)$
and $\Phi(r,\theta,\varphi)$. 
These two profile functions and $P(r,\theta,\varphi)$ describing the
spatial extension of the pion fields will depend on all three (radial,
polar and azimuthal) variables~\footnote{In the present work we
  perform all calculations in the spherical coordinate system.}.  
Since we choose the magnetic field along the $z$ direction, 
we have an axial symmetry, so the profile functions $P$ and $\Theta$
become independent of the azimuthal angle $\varphi$,
and the third profile function $\Phi$ can be selected as $\varphi$.
Thus, one has the following axially symmetric ansatz
\begin{align}
P=P(r,\theta),\quad \Theta=\Theta(r,\theta),\quad
\Phi=\varphi
\label{proffs}
\end{align}
which will be used in the present work.
\section{Classical soliton mass and
  parametrizations of  profile functions}
\label{sec:clmass}

Using the configuration given in Eqs.~(\ref{ansatz})-(\ref{proffs}),  
one can find the mass of the static soliton $M$ in the presence of the 
static magnetic field $\bm{B}_M$ along the $z$ direction. 
The mass functional $M[P,\Theta]$ is explicitly written 
by Eq.~(\ref{eq:Mclass}) in Appendix~\ref{app:mass_momin}.  The field
equations of the soliton can be derived by variation of $M$ with
respect to $P$ and $\Theta$. Since their expressions are rather
lengthy and will not be used here, we will not present them in this
work.  In fact, they are coupled second-order partial differential
equations of the following type\footnote{For the definitions of
  $F_r$, $\Theta_\theta$ etc., see Appendix~\ref{app:mass_momin}. }
\begin{align}
&g(P_{rr},P_{\theta\theta},
 P_{r},P_{\theta},\Theta_r,\Theta_\theta,P,\Theta)=0,\cr
&h(\Theta_{rr},\Theta_{\theta\theta},\Theta_r,\Theta_\theta,
 P_{r},P_{\theta},\Theta,P)=0,\nonumber
\end{align}
and the boundary conditions are determined by the baryon number, i.e.
$B=1$ in the present work. The baryon number of the axially deformed
hedgehog configuration is given by the following expression
\begin{equation}
B = -\frac{1}{\pi}\int\limits_0^{\infty}dr\int\limits_{0}^{\pi}
d\theta\,(P_{r}\Theta_\theta-P_\theta\Theta_r)\sin^2P
{\sin \Theta}.
\label{baryonnumber}
\end{equation}
Since we will use the variational method developed in
Ref.~\cite{Meissner:2008mr}, we will not write the explicit expression
of the solitonic field equations, as mentioned previously. 
This will simplify all unnecessary technical complexities.

However, in order to clarify the form of trial functions to be used
for a minimization process, let us for the moment ignore the
nonspherical deformation effects and assume that the soliton has a
spherical form even if it is affected by the magnetic field. Then the
equation of motion becomes an ordinary but nonlinear differential  
equation. For our purpose, we will rather concentrate on its
linear approximation ($r\rightarrow \infty$) that yields the following
form 
\begin{align}
&P''(r)+\frac{2}{r}P'(r)-\frac{2}{r^2}P(r)\cr
-&\left(m_\pi^2+\frac{2q_eB_M}{3}\right)\!P(r)
-\frac{2(q_eB_Mr)^2}{15}
P(r)=0.
\label{eq:lineq}
\end{align}
Note that the last two terms contribute differently, depending on whether
the magnetic field is strong or weak. They will bring about
interesting consequences and will play a key role in understanding the
present results later. 

In general, Eq.~(\ref{eq:lineq}) has a gaussian 
form of the solution
\begin{align}
P(r)&\sim\frac{1}{2^{1/4}r^2}\exp\Big\{-\frac{q_eB_Mr^2}
{\sqrt{30}}\Big\}\\
\times & U\Big(\frac{-3+\sqrt{30}}{12}+\frac{\sqrt{30}m_\pi^2}
{8q_eB_M},-\frac{1}{2}; \sqrt{\frac{2}{15}}q_eB_Mr^2\Big),
\nonumber
\end{align}
where $U(a,b;c)$ is the confluent hypergeometric function 
of the second type. However, if $m_\pi^2\gg q_eB_M$, then one can
ignore the quadratic term in $B_M$ of Eq.~(\ref{eq:lineq}),
keeping in mind that the soliton is localized at the finite 
region, i.e. even if $r$ is large, the last term  in
Eq.~(\ref{eq:lineq}) is not important due to the localization of
solution.  Then the corresponding solution takes the Yukawa-type form   
\begin{align}
P(r)\sim \frac{1+Ar}{r^2}\,e^{-Ar},\quad
A=\big(m_\pi ^2+\frac{2}{3}q_eB_M\big)^{1/2}.
\end{align}
We will return to the consequences arising from these two different 
behaviors of the solutions, when we discuss the results. 
Having analyzed the characteristics of the solutions at this stage, we
are able to choose the most appropriate forms of the trial profile
functions $P$ and $\Theta$.

As a result, we can apply the following approximations for the
spheroidal solutions  
\begin{align}
P(r,\theta)&=2\arctan\left\{\frac{r_0^2}{r^2}(1+
Ar)[1+u(\theta)]\right\}\cr
&\times \exp\left\{-\beta_0A r-\beta_1q_eB_M r^2)\right\},
\label{Pfun}\\
\Theta(r,\theta)&=\theta+\zeta(r,\theta)\,,
\label{app1}
\end{align}
where $r_0$, $\beta_0$ and $\beta_1$ are variational 
parameters.\footnote{We note that the parametrization in
Eq.~(12) and~(13) are done in a most general form and 
indicates the different field regimes during our variational 
calculations in a natural way.} 
The functions $u$ and $\zeta$ satisfy the inequalities 
$|u|<1$ and $|\zeta|<1$ in the regions  $r\in[0,\infty)$ and
$\theta\in[0,\pi]$.  Thus, the trial function in Eq.~(\ref{Pfun})
correctly reproduces the asymptotic forms of the solutions for both
the weak and strong magnetic fields, and provides the
smooth transitions between these two different cases. 
Furthermore, following the ideas of Ref.~\cite{Yakhshiev:2001ht}, we 
use for the function $u$ the following parametrization
\begin{align}
u(\theta)= q_eB_M
\sum_{n=1}^\infty\gamma_n\cos^n\theta\,,
\label{ufunc}
\end{align}
where the set $\{\gamma_n\}$ consists of variational
parameters in addition to those three mentioned previously. In the  
parametrization of Eq.~(\ref{ufunc}), the cosine functions are chosen  
to maintain the periodicity in $\theta$. Similarly, $\zeta$ can be
selected as 
\begin{align}
\zeta(r,\theta)= q_eB_M re^{-\delta_0^2 r^2}
  \sum_{n=1}^\infty\delta_n\sin (2n\theta)\,, 
\label{zetafunc}
\end{align}
where the set $\{ \delta_n\}$ contains the remaining part 
of all the variational parameters in the present work.  The prefactor 
`$q_eB_M$'  in Eqs.~(\ref{ufunc}) and~(\ref{zetafunc}) is introduced 
from the proper limiting consideration and will smooth the
variational process. 

Note that the arguments of the sine functions in Eq.~(\ref{zetafunc})
are picked out to be a multiple of $2\theta$ in order 
to avoid singularities given in the form `$\sin\Theta/\sin\theta$',
which can be found in the mass functional $M[P,\Theta]$.
Furthermore, the $r$ dependence of $\zeta$ is singled out 
such that the equalities $\Theta(0,\theta)=\theta$ and
$\Theta(\infty,\theta)=\theta$ are reproduced correctly.
The mass functional will be easily extremized in terms of the trial
functions given in Eqs.~(\ref{Pfun})-(\ref{zetafunc}), and  $B=1$  
condition will be naturally satisfied. 

We want to mention that, in order to keep the minimization process
with high accuracy, it is enough to consider only few terms in 
the trial functions~(\ref{ufunc}) and~(\ref{zetafunc}).
Furthermore, the current situation is in a more symmetric level 
than the case in which the nucleon is located in a finite nucleus at a 
given distance from its center. More specifically, when the nucleon
is located inside the finite nucleus, the values 
of the profile functions $P$ and $\Theta$ with the polar angle
given in $\theta\in[0,\pi/2]$ are different from those with
$\theta\in[\pi/2,\pi]$. This is due to the fact that
the external field, which is expressed by the
density distribution function of the external system, depends on 
the radial distance from the center of the nucleus (see
Ref.~\cite{Meissner:2008mr}).  On the other hand, the present case is
symmetric under the change of the polar angles from
$\theta\in[0,\pi/2]$ to $\theta\in[\pi/2,\pi]$, because the external
magnetic field is homogeneously exerted along the $z$
direction. Therefore, the symmetry in the polar angle 
brings about $\gamma_{2n-1}=0~(n=1,2,\dots)$ among 
$\{\gamma_n\}$ in Eq.\,(\ref{ufunc}). 

We perform the variational calculation by minimizing the complete 
energy functional given in Eqs.\,(\ref{eq:Mclass}) 
and\,(\ref{eq:deltaMclass}), using the trial profile functions 
given in Eq.\,(\ref{Pfun}) and \eqref{app1}. This 
approach is rather accurate, because 
both the solutions near the origin ($r\rightarrow 0$) and asymptotic
region ($r\rightarrow\infty$) are properly given. 
The variational parameters introduced above connect smoothly 
the solution near the origin with the asymptotic one, which
reproduces almost the exact solutions.  
For example, in the case of a free nucleon, we obtain almost the same
results by either using the variational approach or directly solving
the differential equations. Both the results differ within 1~\% 
(e.g. see Table~1 of Ref.~\cite{Meissner:2008mr}).
In the present work, the same level of high accuracy is achieved.

 \begin{figure}[htp]
\begin{centering}
\includegraphics[scale=0.85]{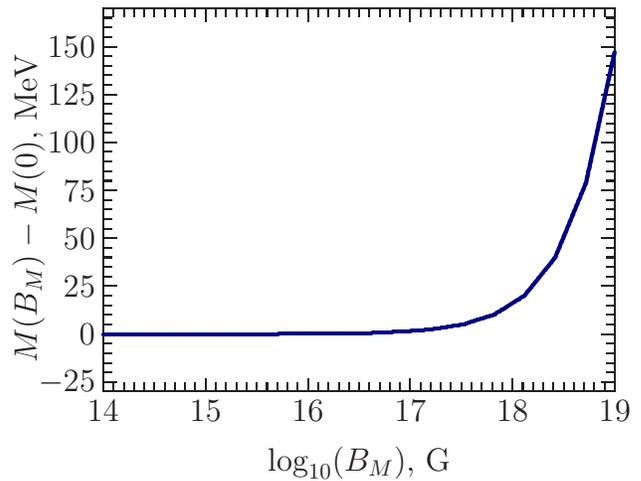}
\end{centering}
\caption{(Color online)
The change of the classical soliton mass  as a function of  
the external magnetic field given in the log scale. $M(B_M)$ and
$M(0)$ denote respectively the values of the mass obtained with and
without the magnetic field exerted.} 
\label{fig:1}
\end{figure}
Figure~\ref{fig:1} draws the results for the change of the classical
soliton mass due to the external magnetic field, i.e. $M(B_M)-M(0)$,
where $M(B_M)$ and $M(0)$ denote respectively the values of the mass
obtained with and without the magnetic field exerted. 
The mass of the classical soliton remains constant till the strength
of the magnetic field reaches around $10^{17}\,\mathrm{G}$.
However, as the magnetic field gets stronger than $10^{17}$~G,  the
value of the soliton mass starts to increase slowly till $B_M\approx
10^{18}$~G.  If one raises the magnitude of the magnetic field,
then the soliton mass starts to rise rather rapidly. When the magnitude
of the magnetic field becomes $10^{19}$ G, the soliton mass acquires
approximately additional 150 MeV by the external magnetic field.  

Before we discuss the main results of the present work, we want to
examine the values of the variational parameters for the profile
functions. In Table~\ref{tab:1}, we list their numerical results 
determined at the several selected values of the magnetic field.  
\begin{table}[hbt]
\caption{Variational parameters for the profile functions $P$ and
  $\Theta$ at some selected values of the external magnetic field
  $B_M$. 
}
\begin{ruledtabular}
\begin{tabular}{l|cccc}
\quad$B_M$ &$0$&$10^{15}$\,G&
$10^{17}$\,G&$10^{19}$\,G\T\\
\hline
$r_0$, fm$^2$ &0.95646& 0.95641& 0.95200 & 0.97324\T\\
$\beta_0$ & 1.31568 & 1.31554& 1.30447 & 0.93320\\
$\beta_1$ & 0 & 0 & 0 &0.21958\\
$\gamma_2$, fm$^{2}$ & 0 & $\!\!\!\!-$0.64430& 0.12305 & 0.33700\\
$\gamma_4$, fm$^{2}$ & 0 &  0.30370& 0.21985 &0.08227\\
$\gamma_6$, fm$^{2}$ & 0 & $\!\!\!\!-$0.10019& $\!\!\!\!-$0.14775&0.21615  \\
$\delta_0$, fm$^{-2}$ & 4.23604 & 3.90049 &2.84256 &3.21149 \\ 
$\delta_1$, fm & 0 & 0.13997 & 0.09016 & 0.9366 \\
$\delta_2$, fm & 0 & 0.24411 & 0.00207 & 0.00174 \\
\end{tabular}
\end{ruledtabular}
\label{tab:1}
\end{table} 
Among the parameters presented in Table~\ref{tab:1}, nonzero 
values of $\gamma_n$'s are responsible for the deviation  
of the $P$ equi-surfaces from the spherical form, whereas
$\delta_n$'s  $(n>0)$ exhibit how the shape of the profile function is
distorted from the spherically symmetric hedgehog form. One can see
from Table~\ref{tab:1} that at $B_M=10^{15}$~G, which characterizes
the strength of the magnetic fields in magnetars, the $P$
equi-surfaces already deviate from the spherically symmetric form.    

It is interesting to observe that the value of $\beta_1$ is almost
intact even at the upper limit of the strength of the magnetic fields
in neutron stars ($\sim 10^{17}$\,G)~\cite{Thompson:1996pe}. 
However, if the strength of the magnetic field gets stronger, then
its value is not zero anymore (see the corresponding value 
listed in the last column of Table~\ref{tab:1} for $B_M=10^{19}$~G). 
In order to understand this behavior, we need to scrutinize the
exponential term in Eq.~(\ref{Pfun}). When $A$ in the first term is
much larger than $(r q_e B_M)^2$ in the second one, for example, when 
$r^2(q_e\times10^{17}\,\mathrm{G})^2   \sim  10\,\mathrm{MeV}^2\ll
(m_\pi^2+2q_eB_M/3) \approx m_\pi^2\sim 0.18\,\mathrm{GeV}^2$
numerically, then the asymptotic solution is not much influenced 
by the second term for the typical soliton size ($r\sim$1
fm). However, when the magnetic field is extremely strong,
i.e. $B_M\sim  10^{19}\,{\rm G}$, we find $q_eB_M>m_\pi^2$.
Thus, the second term dominates over the first one. 
It implies that the profile $P(r,\theta)$ will be shrunken by the
second Gaussian term and $\beta_1$ is not zero anymore.

 \begin{figure*}[htb]
\begin{centering}
\includegraphics[scale=0.85]{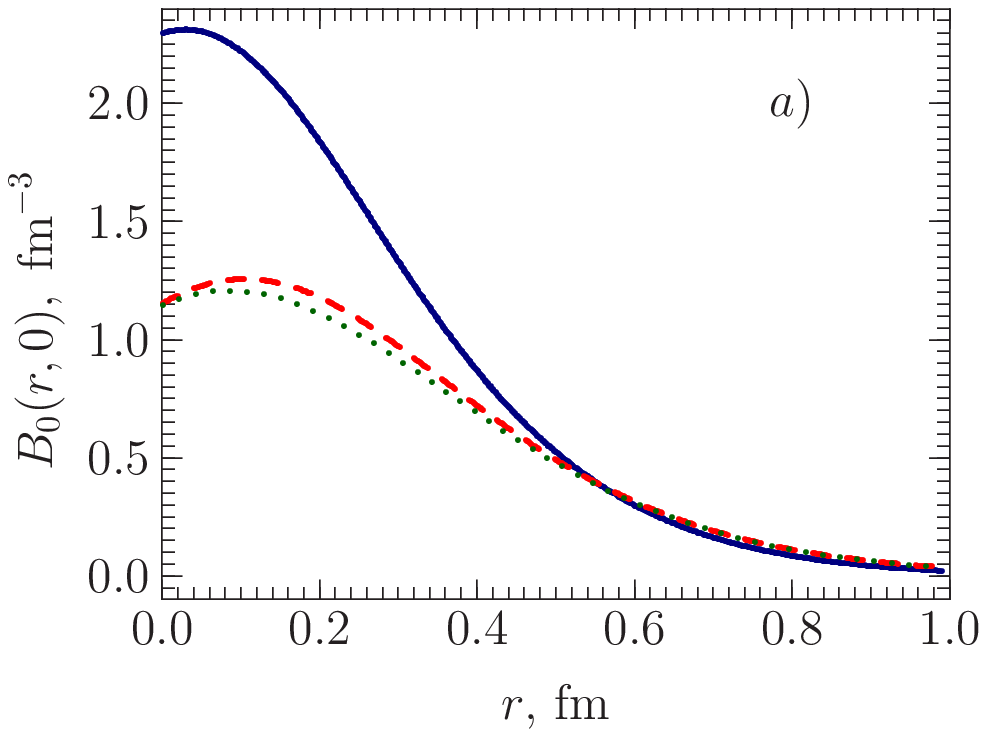}\hskip 0.35cm
\includegraphics[scale=0.85]{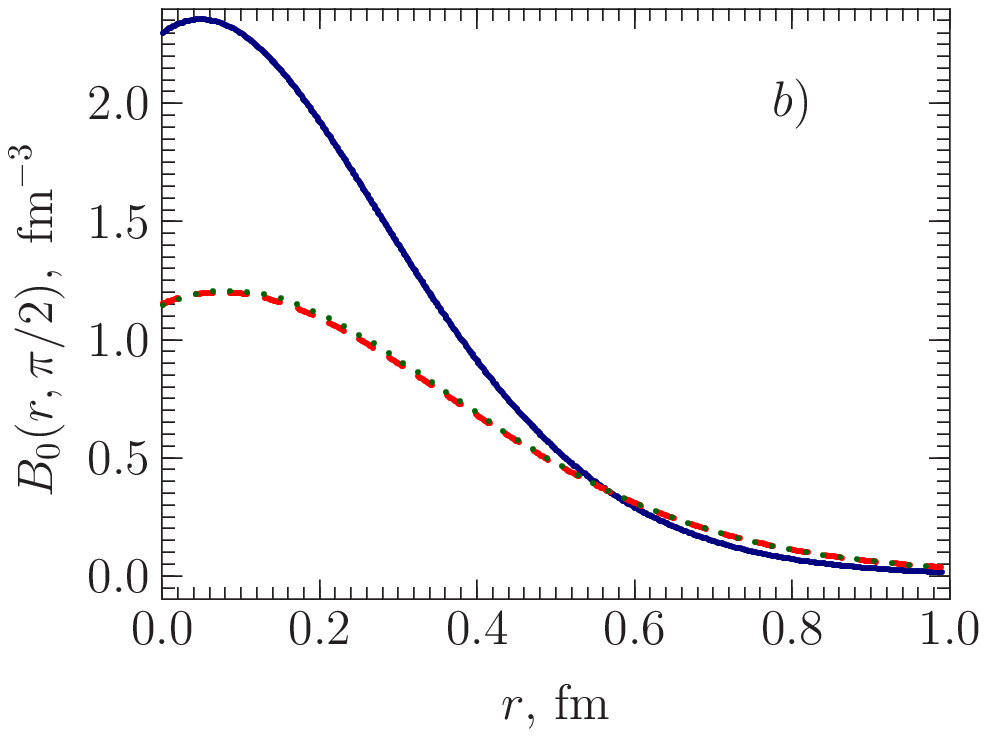}
\end{centering}
\caption{(Color online)
The baryon charge distributions along the $z$ direction as functions
of $r$ (in the left panel denoted by a)) and those in the 
perpendicular plane to the $z$ axis as functions of $r$ (in the right
panel denoted by 
b)), respectively. The solid curves depict the results with   
$B_M=10^{19}$\,G, the dashed ones draw those with $B_M=10^{17}$\,G,  
and the dotted ones correspond to the case of $B_M=0$, respectively.} 
\label{fig:2}
\end{figure*}
To understand the above-mentioned nature more clearly, we will delve
into the baryon charge distribution of the axially deformed skyrmion,
which is expressed as\footnote{See Eq.~(\ref{baryonnumber}) above.} 
\begin{align}
B_0(r,\theta) = -\frac{P_{r}\Theta_\theta-P_\theta\Theta_r}{2\pi^2r^2} 
\left(\frac{\sin \Theta}{\sin\theta}\right)\sin^2P.
\label{baryonnumberdensity}
\end{align}
It will explicitly reveal how the soliton undergoes deformation in the
presence of the strong magnetic field.  In the 
Fig.~\ref{fig:2}a, we depict the profiles of the baryon charge
distributions along the $z$ direction ($\theta=0$), 
while the Fig.~\ref{fig:2}b draws those in the
perpendicular plane to the $z$ axis ($\theta=\pi/2$).
Dotted curves correspond to the results with $B_M=0$, which should be
spherically symmetric and are the same in both the left and right
panels. We can take them as a reference for comparison.  
Taking the value of the magnetic field to be $B_M=10^{17}~\mathrm{G}$,
we see that the charge distribution of the soliton along the $z$
direction is deformed slightly, whereas it remains the same as that in
the absence of magnetic field as shown in the right panel.
If we take $B_M=10^{19}$\,G, which can be realized in URHICs at the
LHC, then the baryon charge distribution displays evidently the
deformation of the soliton both along  the $z$ direction and in the
perpendicular plane to it. 

\begin{figure*}[htp]
\begin{centering}
\includegraphics[scale=0.85]{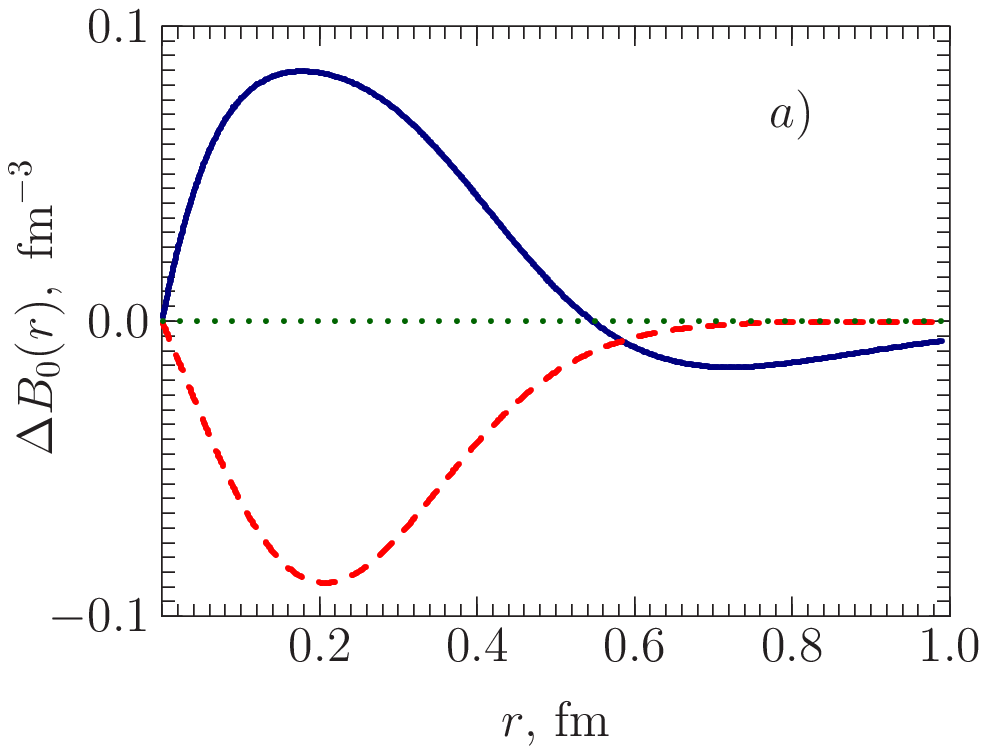}\hskip 0.35cm
\includegraphics[scale=0.85]{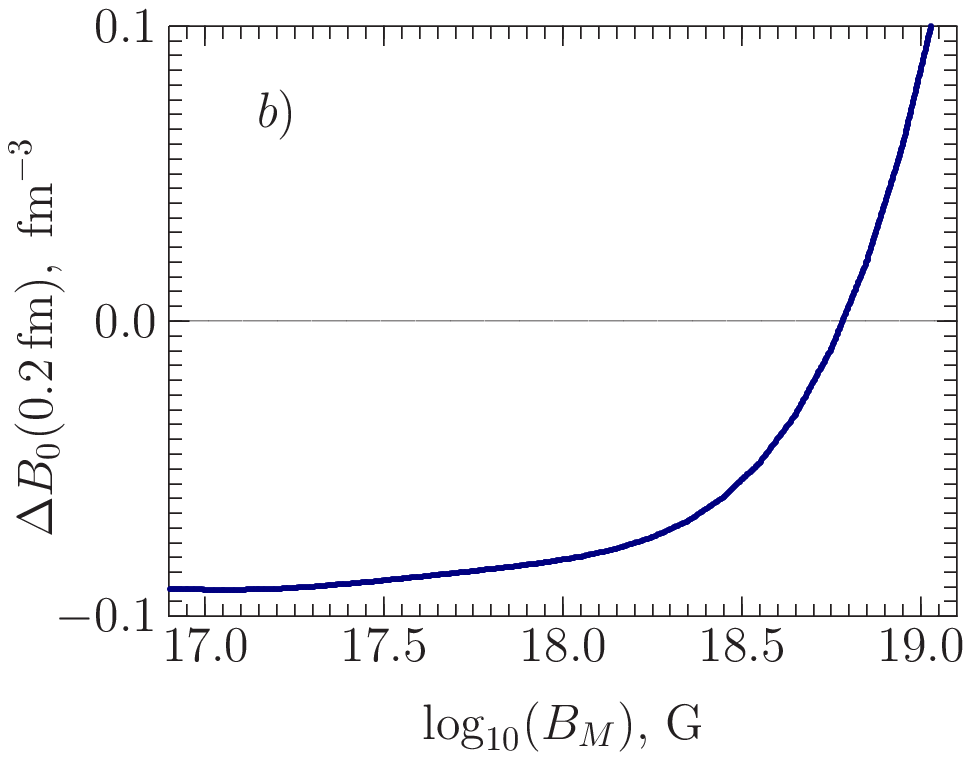}
\end{centering}
\caption{(Color online)
In the left panel denoted by a), the results of the anisotropy $\Delta
B_0(r)$ defined in Eq.~\eqref{eq:deltaB} as functions of
$r$ and in the right panel denoted by b) that of $\Delta B_0(0.2\,{\rm 
  fm})$ fixed at $r=0.2\,\mathrm{fm}$ as a function of the magnetic
field. Notations are the same as in
Fig.~\ref{fig:2}.}  
\label{fig:3}
\end{figure*}

To illuminate how the baryon charge distribution undergoes the change
as the strength of the magnetic field is varied, we define the 
\emph{anisotropy of the baryon charge distribution} as
\begin{align}
\Delta B_0(r)\equiv B_0(r,{\pi}/{2})-B_0(r,0)
\label{eq:deltaB},
\end{align}
where $B_0(r,{\pi}/{2})$ represents the baryon charge distribution in
the perpendicular plane to the $z$ axis, and $B_0(r,0)$ denotes that
along the $z$ direction. 
Equation~\eqref{eq:deltaB} shows how the isotropy of the baryon charge
distribution is broken by the magnetic field. In the Fig.~\ref{fig:3}a we
illustrate the results of $\Delta B_0$ as functions 
of $r$ and in the Fig.~\ref{fig:3}b those at $r=0.2\,\mathrm{fm}$ as a
function of the magnetic field, respectively. 
The results of Fig.~\ref{fig:3}a clearly show that
when $B_M=10^{17}$\,G, which corresponds to the dashed curve, the
soliton is more deformed along the $z$ direction than in the
perpendicular plane to it. Moreover, it mainly occurs in the
core part of the soliton. It implies that the baryon charge distribution 
will be taken slightly as a cigar-type form, since the results of $\Delta
B_0(r)$ decreases in the core part. If one takes the stronger value of 
the magnetic field, i.e. $B_M=10^{19}$ G, then the baryon charge
distribution is drastically changed from the previous case of
$B_M=10^{17}$ G. The core part of the soliton undergoes the
deformation in the $xy$ plane more strongly than along the $z$
direction. On the other hand, when it comes to its peripheral part,
the situation is  reversed. That is, while the peripheral shape of the
soliton is less distorted than in the core part, the soliton is
slightly more deformed along the $z$ direction in comparison with that
in the perpendicular plane to it.

To see the process of the soliton deformation more closely, we
scrutinize $\Delta B_0$ at a fixed value of $r$, for example, at
$r=0.2$~fm, as the $B_M$ field varied from $10^{17}$~G to
$10^{19}$~G. The corresponding result is illustrated in the 
 Fig.~\ref{fig:3}b. When the strength of the magnetic field is
given between $10^{17}$~G and $10^{18}$~G, we can clearly observe that
the core part of the soliton is more deformed in the $xy$ plane,
compared with that along the $z$ direction. However, if we further
increase the strength of the magnetic field close to $10^{19}$~G, the
situation becomes other way around, i.e. the core part of the soliton
is deformed more strongly along the $z$ direction in comparison with
that in the perpendicular plane to it. 

In general, the baryon charge distribution is more compactly deformed 
in the presence of the strong magnetic field. This can be observed
by comparing the solid curves with dotted ones in the
Fig.~\ref{fig:2}. Since the quadratic term with regards to $B_M$ in
Eq.~\eqref{Pfun} come into dominant play when the magnetic field is
very strong. In fact, this is related to the quadratic term like a
harmonic oscillator potential in the approximated differential
equation given in Eq.~\eqref{eq:lineq} in the asymptotic limit, which
plays effectively a role of a confining potential that arises
from the the strong magnetic field. The physical implications of this
confining potential are that the pions are localized and are
forced to be confined by the external strong magnetic 
field.

Here, it is necessary to remind that the baryon is a topological object 
made of the nonlinearly interacting pions.
In this context, although the Skyrme model has no explicit 
quark degrees of freedom, the obvious charged pion localizations due
to the external magnetic field will localize also the neutral pions by
means of the nonlinear interactions. Moreover, the quantization by
rotation in isospin space infers that both the charged and neutral
pions are under the influence of the strong magnetic field. 
It can be explicitly seen from the expressions of the baryon charge  
distribution~(\ref{baryonnumberdensity}) and the mass
functional (see Eqs.\,(\ref{eq:Mclass}) 
and\,(\ref{eq:deltaMclass})), which 
do not distinguish the charged components of the pion 
fields.\footnote{There is yet another effect of the 
magnetic field on the quarks inside the neutral pion 
trough the wave function deformations. This effect comes from 
higher-order corrections with respect to the external magnetic field,
which is not considered in this work. }

We can examine the corresponding localizations of the nonlinearly
interacting pions by considering the baryon charge
distribution within a certain region. For example, we integrate
the baryon charge distribution up to 1 fm 
\begin{align}
B_{\rm (1\,fm)}=
\int\limits_0^{1\,{\rm fm}} 
r^2dr\int d\Omega\, B_0
\label{B1fm}
\end{align}
with the magnetic field varied. Then, comparing the results with
different values of $B_M$, we can see how the 
charged pions are forced toward the core region inside a
nucleon. Taking three different values of $B_M$, we obtain the
following results: 
$B_{\rm  (1\,fm)}=0.9014$ for $B_M=0$,  
$B_{\rm (1\,fm)}=0.9024$ for $B_M=10^{17}$\,G
and $B_{\rm  (1\,fm)}=0.9665$ for $B_M=10^{19}$\,G, 
respectively\footnote{Of course, we get $B=1$ in all cases if one
  integrates properly all over the region, as it should be.}. 
The comparison of these values indicates that the baryon charge
distribution is indeed squeezed into the core region
due to the localization of the charged pions.

\section{Quantization of the spheroidal soliton} 
\label{sec:quant}
We are now in a position to discuss the quantization of the axially
deformed soliton, i.e., the spheroidal one and the relevant results. The
quantization of a spherically symmetric chiral soliton is generally
performed by introducing the zero-mode quantization with the
collective coordinates introduced~\cite{Adkins:1983ya}.  As we already
discussed in the previous Section, the spherical symmetry of the
soliton is already broken in the presence of the magnetic
field. However, we still have an axial symmetry as presented in
Eqs.~(\ref{ansatz})-(\ref{proffs}). Thus, we consider  
independent rotations in the coordinate and isospin spaces as follows 
\begin{align}
P=P\left({\cal R}^{-1}(t)\bm{r}\right), \quad  
\bm{N}={\cal I}(t){\bm{N}}\left({\cal R}^{-1}(t)\bm{r}\right),
\end{align}
where ${\cal R}$ and ${\cal I}$ represent the SO (3) rotational and 
iso-rotational matrices, respectively. Having carried out these slow
time-dependent rotations and performed the spatial integration, we
arrive at a collective Lagrangian 
\begin{align}
L=&-M
+\frac{\omega_1^2\!+\!\omega_2^2}{2}\Lambda_{\omega\omega,12} -
    (\omega_1\Omega_1\!+\!\omega_2\Omega_2)
    \Lambda_{\omega\Omega,12}\cr 
  & + \frac{\Omega_1^2\!+\!\Omega_2^2}{2}
    \Lambda_{\Omega\Omega,12}+\frac{(\omega_3\!-\!\Omega_3)^2}{2}
    \Lambda_{\omega\Omega,33}.
\label{Lag-t} 
\end{align}
Here $\omega_i$ and  $\Omega_i$  denote the angular velocities in
isospin and coordinate spaces, respectively. The explicit  
expressions of the functionals  $\Lambda[P,\Theta]$ 
can be found in Appendix~\ref{app:mass_momin}. 

Defining the canonical conjugate variables in the body-fixed reference
system  as 
\begin{eqnarray}
T_i=\frac{\partial{L}}{\partial\omega_i}\mbox{\quad and\quad}
J_i=\frac{\partial{L}}{\partial\Omega_i}\,,
\end{eqnarray}
we derive from the time-dependent Lagrangian in Eq.~\eqref{Lag-t}  the
collective Hamiltonian as 
\begin{align}
\hat H&=M+\frac{\hat T_3^2}{2\Lambda_{\omega\Omega,33}}
+\frac{(\hat T_1\hat J_1+\hat T_2\hat J_2)\Lambda_{\omega\Omega,12}}
{\Lambda_{\omega\omega,12}\Lambda_{\Omega\Omega,12}
-\Lambda_{\omega\Omega,12}^{2}}\cr
&+\frac{(\hat T_1^2+\hat T_2^2)\Lambda_{\Omega\Omega,12}+
(\hat J_1^2+\hat J_2^2)\Lambda_{\omega\omega,12}}
{2(\Lambda_{\omega\omega,12}\Lambda_{\Omega\Omega,12}
-\Lambda_{\omega\Omega,12}^{2})}.
\label{Ham-final}
\end{align}
Diagonalizing the Hamiltonian of Eq.~\eqref{Ham-final}, we obtain the 
baryon eigenstates $|T,T_3;J,J_3\rangle$ and the energies of the
axially deformed nucleon and the $\Delta$ isobar:   
\begin{align}
\label{E-final}
E&=M+\frac{T_3^2}{2\Lambda_{\omega\Omega,33}}\\
&+\frac{\Lambda_{\Omega\Omega,12}+\Lambda_{\omega\omega,12}
-2\Lambda_{\omega\Omega,12}}
{2(\Lambda_{\omega\omega,12}\Lambda_{\Omega\Omega,12}
-\Lambda_{\omega\Omega,12}^{2})}\big(T(T+1)-T_3^2\big).\nonumber
\end{align}
From the third term of Eq.~\eqref{E-final}, one observes that in the
presence of the external magnetic field the degeneracy in the energy
between the different isospin states of the $\Delta$ isobar are
partially lifted. For example, the proton and neutrons are still in 
degeneracy, i.e. $m_{\rm p}=m_{\rm n}$, while  the $\Delta$ isobar
isospin states are partially split, i.e.
$m_{\Delta^{++}}=m_{\Delta^-}\neq m_{\Delta^+}=m_{\Delta^0}$.

The results for the masses of baryons at certain values of the
magnetic field $B_M$ are listed in Table~\ref{table:2}.
\begin{table}[hbt]
\caption{Masses of baryons
at some selected values of the external magnetic field $B_M$.
All masses are given in units of MeV.
}
\begin{ruledtabular}
\begin{tabular}{c|cccc}
$B_M$ &$0$&$10^{15}$\,G&$10^{17}$\,G&$10^{19}$\,G\T\\
\hline
$m_{\rm n,p}$
&\,\,939.8035&\,\,939.8212&\,\,\,941.5769 &\,1113.4133 \T\\
$m_{\Delta^{++},\Delta^{-}}$ 
& 1233.6770 & 1233.6951 & 1236.5624&1530.4224\\
$m_{\Delta^{+},\Delta^0}$ 
& 1233.6770 & 1233.6949 & 1236.3618& 1507.7573\\
\end{tabular}
\end{ruledtabular}
\label{table:2}
\end{table}
As in the case of the classical soliton, the masses of the nucleons and
$\Delta$ isobars are almost intact till the strength of the magnetic
field is reached at around $10^{17}$~G. Keeping in mind that the
magnetic field in magnetars is approximately $B_M=10^{15}$\,G, the
baryon masses are almost not changed. However, if one further
increases the strength of $B_M$, the masses of all the nucleons and
$\Delta$ isobars start to grow. At $B_M=10^{17}$\,G the change of the
baryon mass is already not negligible.  Then, when it is reached to 
$B_M=10^{19}$~G, the masses increase by about $15-20$~\%.
Note that $\Delta$-isobar states actually remain degenerate  
even though the magnetic field gets very strong.
Only at very large values of the magnetic field, the degeneracy of the
$\Delta$ isobars will be partially lifted as discussed above. 

\begin{figure}[htp]
\begin{centering}
\includegraphics[scale=0.8]{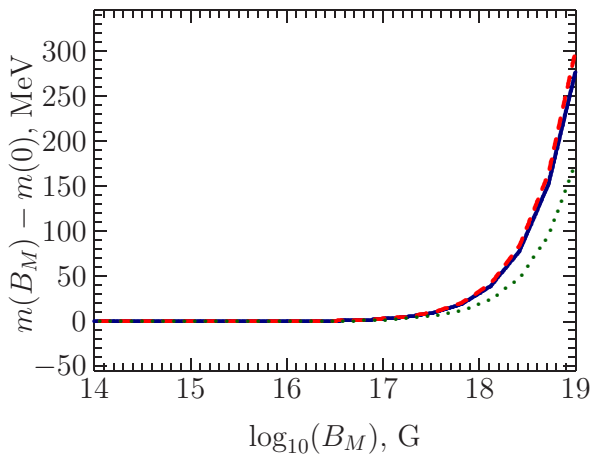}
\end{centering}
\caption{(Color online)
The changes of the baryon masses as a function of the magnetic
field. The solid curve depicts $m_{\Delta^0}$, whereas the dashed one
draws $m_{\Delta^-}$. The dotted one represents $m_{\rm n}$,
respectively.} 
\label{fig:4}
\end{figure}

In Fig.~\ref{fig:4} we show how the masses of the baryons will be
changed as $B_M$ increases.
The results look very similar to the change of the classical soliton
mass as shown in Fig.~\ref{fig:1}. However, the rates of the increment
in the masses of the baryons are still different. The reason can be
found in the changes of the moments of inertia\footnote{The formula
  for the moments of inertia, see Eq.~\eqref{eq:A5}.}. Since the
soliton is deformed in the presence of the strong magnetic fields, the
magnitudes of the moments of inertia are decreased. 
It indicates that not only the baryon charge distribution is
changed but also the mass distribution inside the soliton becomes 
more compact in the presence of the strong magnetic field than in free
space. As was done in the case of the baryon charge distribution, we
can consider the integrate value of the mass distribution up to $1$~fm
(see Eq.\,\eqref{B1fm}). Then we obtain the results at three different
values of $B_M$ as follows: 
$M_{\rm (1\,fm)}= 0.818M$ for $B_M=0$, 
$M_{\rm (1\,fm)}= 0.820M$ for $B_M=10^{17}$\,G and 
$M_{\rm (1\,fm)}= 0.911M$ for $B_M=10^{19}$\,G. 
This indicates that the masses of the baryons tend to be more compact
in the presence of the magnetic fields than in free space. 

\begin{figure}[htp]
\begin{centering}
\includegraphics[scale=0.8]{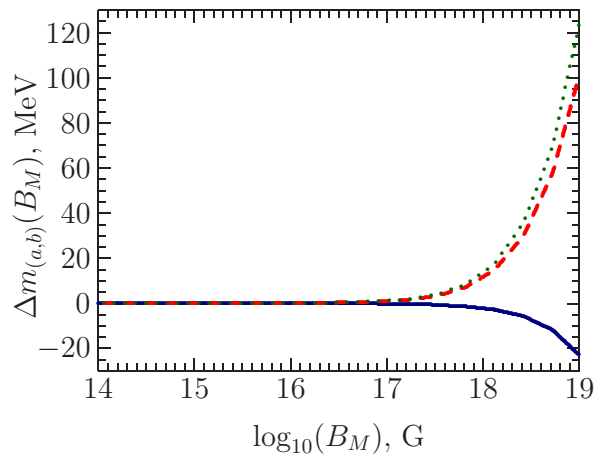}
\end{centering}
\caption{(Color online)
The change of the baryon mass splittings in the presence of the
magnetic field.  The solid curve draws the result of $\Delta
m_{(0,-)}(B_M)$, whereas the dashed one depicts 
$\Delta m_{\rm (0,n)(B_M)}$. The dotted one shows $\Delta m_{\rm
  (-,n)}(B_M)$. For the definitions of $\Delta m_{(a,b)}$, see
Eqs.~(\ref{DeltaM0n})-(\ref{DeltaMmn}).
  }
\label{fig:5}
\end{figure}

It is also very interesting to examine the moments
of inertia for the spheroidal solitons. We first define the following
quantities 
\begin{eqnarray}
\Delta m_{(0,-)} (B_M)&=&
[m_{\Delta^{0}}(B_M)-m_{\Delta^{-}}(B_M)]\cr
&-&[m_{\Delta^{0}}(0)-m_{\Delta^{-}}(0)],
\label{DeltaMm0}
\\
\Delta m_{\rm (0,n)} (B_M)&=&
[m_{\Delta^{0}}(B_M)-m_{\rm n}(B_M)]\cr
&-&[m_{\Delta^{0}}(0)-m_{\rm n}(0)],
\label{DeltaM0n}
\\
\Delta m_{\rm (-,n)} (B_M)&=&
[m_{\Delta^{-}}(B_M)-m_{\rm n}(B_M)]\cr
&-&[m_{\Delta^{-}}(0)-m_{\rm n}(0)].
\label{DeltaMmn}
\end{eqnarray}
They describe how much the mass splittings of the baryons undergo the
changes in the presence of the magnetic field. The results are
illustrated in Fig.~\ref{fig:5}.  
Here we explicitly demonstrate that the moments of inertia  
decrease, which bring about the rise of the $\Delta-N$ mass
splittings, which are illustrated in the dashed and dotted curves in
Fig.~\ref{fig:5}. One can also observe that the mass degeneracy in the
different isospin states of the $\Delta$ isobars is lifted, as shown
in the solid curve of Fig.~\ref{fig:5} . While the degeneracy
is more or less kept to be intact till $B_M=10^{17}$\,G, it starts to be
removed. If $B_M$ continues to increase, the splitting between the
${\Delta^{0}}$ and ${\Delta^-}$ masses becomes prominent.

Finally, we want to mention that there is still a caveat that is
related to the strong magnetic fields.  
A novel feature emerges when the magnetic field is very strong, called
the Paschen-Back (PB) effect~\cite{Paschen:1921}. Originally, the PB
effect arises when the strength of the magnetic field dominates over
the spin-orbit coupling of an atomic system. In the presence of the
weak magnetic field, all the eigenstates of an atom are
split, which is known as the anomalous Zeeman effect. However, if the
magnetic field is so strong that it overcomes the spin-orbit
interaction, then the spherical symmetry is completely 
broken, so that the total angular momentum squared, $J^2$, is no more
a good quantum number but $L_z$ and $S_z$ are the good quantum
numbers~\cite{Sakurai:2011zz}.  However, we still have cylindrical
symmetry or axial symmetry in the presence of the constant external
magnetic field along a specific direction as discussed in this
work. Thus, $2(2l+1)$ degeneracy in $m_l+m_s$ will appear. This is
called the PB effect. In fact, Iwasaki et al. discussed the PB
effect~\cite{Iwasaki:2018pby}, when the strong magnetic field ($\sim
10^{19}$~G) is exerted on a charmonium system. They found a very
interesting feature: The strong magnetic field induces mixing between
$S=0$ and $S=1$ states. This may lead to the mixing between the
$\eta_c$ and $J/\psi$ in $S=1$ and $S_z=0$ states. It implies that
when the magnetic field is very strong, one can expect the same
phenomena in a baryonic system such as the mixing between the proton
with $S=1/2$ and $S_z=1/2$ and the $\Delta^+$ isobar with $S=3/2$ and
$S_z=1/2$. We will investigate this important physics elsewhere. 

\section{Summary and Outlook}
\label{sec:summary&outlook}

In the present work, we investigated how the nucleons and 
$\Delta$ isobars undergo the deformation in the presence of the strong
magnetic field within the framework of the Skyrme model. We first
examined the changes of the classical soliton under the influence of
the strong magnetic field. The mass of the classical 
soliton remains unchanged till the magnitude of the magnetic field
reached $10^{17}$~G. However, if the magnetic field gets stronger than
this value, the mass starts to increase. The soliton is deformed in a 
rather nontrivial way as the strength of the magnetic field varied. We 
exhibited explicitly and thoroughly how the soliton 
properties were changed as the magnetic field was altered. When the
magnitude of the magnetic field 
is $10^{17}$~G, the soliton was deformed more strongly along the $z$
direction than in the perpendicular plane to it. The core
part of the soliton was mainly modified, which indicates that the
shape of the soliton will turn to a cigar-type form. If
the value magnetic field was taken to be  $10^{19}$~G, then the baryon
charge distribution was drastically altered. The core part of the soliton
was deformed more strongly in the $xy$ plane than along the $z$
direction. On the contrary, the peripheral shape of the soliton was
less distorted than in the core part, whereas the soliton was 
slightly more deformed along the $z$ direction than in the
perpendicular plane to it.

We performed the zero-mode quantization of the spheroidal soliton in
the presence of the magnetic field. We found that the solitonic
moments of inertia decreases as the magnetic field increases. 
It means that the masses of the nucleon and $\Delta$ isobar should
get larger. Moreover, we observed that $\Delta$-N mass splitting also
increases. The spherical nucleon in free space was deformed into a
cigar-type form when the magnetic field was present. The case of the
$\Delta$ isobars was similar to the nucleon case but their masses
increased slightly more than the nucleon did as the magnetic field is
strengthened. We found that the mass of $\Delta^{++}$
is degenerate with that of $\Delta^-$, whereas $\Delta^+$ has the same
mass as $\Delta^0$. However, the mass degeneracy was partially
lifted. 

From the present work, we conclude that there is no need to consider
the effects of the strong magnetic field in analyzing
the equation of the states (EoS) at high densities that may 
exist in interiors of compact stellar objects, since the nucleon
masses are almost intact till the magnitude of the magnetic field
reaches $10^{17}$~G. 
It is interesting to see that one can make the similar conclusion 
from the recent studies on the EoS of strongly magnetized 
quark matter within the Nambu-Jona-Lasinio 
model~\cite{Avancini:2017gck}.  
However, when it comes to ultra-relativistic
heavy ion collisions at the LHC, it is of great significance to take
into accounts the effects coming from the strong magnetic field. 
This will lead to nontrivial consequences. Furthermore,
generalizations of the model may be performed by including the
explicit isospin breaking effects  in the mesonic sector in order to
study the changes  in the neutron-proton mass difference under the
influence of the external magnetic field. The Paschen-Beck effects on
baryonic systems are yet another interesting issue, which can be
investigated as future works. The relevant works are under way.

\section*{Acknowledgments}
We are grateful to P. Gubler, A. Hosaka, K. Itakura, T. Maruyama for
useful discussions.  U.Y. and H.-Ch.K. would like to express their  
gratitude to the members of the Advanced Science Research Center at
Japan Atomic Energy Agency for the hospitality, where the present work
was done. This work is supported by the Basic Science Research Program
through the National Research Foundation (NRF) of Korea funded by the
Korean government (Ministry of Education, Science and Technology,
MEST), No.~2016R1D1A1B03935053 (UY) and
No.2018R1A2B2001752~(HChK).  

\begin{appendix}
  \section{Mass and moments of inertia of the 
    spheroidal soliton}   
\label{app:mass_momin}
\def\theequation{\Alph{section}.\arabic{equation}}
\setcounter{equation}{0}

For convenience, we introduce the following short-handed notations:
\begin{align}
&P_r\equiv\partial_r P,~P_\theta\equiv\partial_\theta P,~
\Theta_r\equiv\partial_r\Theta,~\Theta_\theta\equiv\partial_\theta\Theta,\cr
&S_P\equiv \sin P,\quad  C_P\equiv \cos P,\quad S_\Theta
\equiv \sin \Theta,\cr
&C_\Theta\equiv \cos \Theta,\quad
s_\theta\equiv\sin\theta,\quad\,\, c_\theta\equiv\cos\theta.
\end{align}
The classical soliton mass $M$ in the Lagrangian in Eq.~(\ref{Lag-t}) 
and the explicit change of the soliton mass $\Delta M$ in the external
magnetic field are expressed as follows:
\begin{align}
M&=\intrT 
\Big\{\frac{F_\pi^2}{4r^2}\Big[P_\theta^2+r^2P_r^2\cr
&+
S_P^2
\Big(\frac{S_\Theta^2}{s_\theta^{2}}+\Theta_\theta^2+r^2\Theta_r^2\Big)
\Big]\cr
&+\frac{S_P^2}{e^2r^4}
\Big[\frac{S_\Theta^2}{s_\theta^{2}}\left(P_\theta^2+r^2P_r^2\right)\cr
&+S_P^2\frac{S_\Theta^2}{s_\theta^{2}}
\left(\Theta_\theta^2+r^2\Theta_r^2\right)
+r^2\left(P_r\Theta_\theta-P_\theta\Theta_r\right)^2\Big]
\cr
&+\frac{m_\pi^2F_\pi^2}{2}(1-C_P)\Big\}+\Delta M,
\label{eq:Mclass}\\
\Delta M&=\intrT 
\Big\{\frac{F_\pi^2}{16}\cr
&+\frac{1}{4e^2r^2}(P_\theta^2
+r^2P_r^2+S_P^2(\Theta_\theta^2+r^2\Theta_r^2)
\Big\}\cr
&\times q_eB_M(4+q_eB_M r^2s_\theta^2)S_P^2S_\Theta^2.
\label{eq:deltaMclass}
\end{align}
The generic form for the moment of inertia is defined as  
\begin{eqnarray}
\Lambda&=&2\intrT\,\,\lambda\,, 
\end{eqnarray}
where the contributions from the different parts of the
Lagrangian~(\ref{Lag-t}) are given as 
\begin{align}
\label{eq:A5}  
\lambda_{\omega\omega,12}&=
\Delta\lambda_{\omega\omega,12}+
\frac{F_\pi^2}8\left(1+C_\Theta^2\right) S_P^2\cr
&+\frac{S_P^2}{2e^2r^2}
\Big[\left(1+C_\Theta^2\right)\left(P_\theta^2+r^2P_r^2\right)
\cr&
+S_P^2
\Big(\frac{S_\Theta^2}{s_\theta^{2}}
+C_\Theta^2\left(\Theta_\theta^2+r^2\Theta_r^2\right)
\Big)\Big],\\
\lambda_{\omega\Omega,12}&=
\Delta \lambda_{\omega\Omega,12}+\frac{F_\pi^2}8
\Big(c_\theta C_\Theta\frac{S_\Theta}{s_\theta}
+\Theta_\theta\Big) S_P^2\cr
&+\frac{S_P^2}{2e^2r^2}
\Big[c_\theta C_\Theta\frac{S_\Theta}{c_\theta}
\Big(P_\theta^2+r^2P_r^2\cr
&+S_P^2\left(\Theta_\theta^2+r^2P_r^2\right)\!\Big)
+S_P^2{S_\Theta^2}{s_\theta^{-2}}\Theta_\theta
\cr
&+r^2P_r(P_r\Theta_\theta-\Theta_rP_\theta)\Big],\\
\lambda_{\Omega\Omega,12}&=
\Delta\lambda_{\Omega\Omega,12}+
\frac{F_\pi^2}8
\Big[P_\theta^2+S_P^2
\Big(c_\theta^2\frac{S_\Theta^2}
{s_\theta^2}+\Theta_\theta^2\Big)\Big]
\cr
&+\frac{S_P^2}{2e^2r^2}
\Big[\frac{S_\Theta^2}{s_\theta^2}\Big((1+c_\theta^2)
(P_\theta^2+S_P^2\Theta_\theta^2)
\cr
&+r^2(P_r^2+S_P^2\Theta_r^2)
c_\theta^2\Big)\cr
&+r^2(P_r\Theta_\theta-P_\theta\Theta_r)^2\Big].
\end{align}
and one can also note, that the moment of inertia corresponding 
to the quantization axis does not depend explicitly on the magnetic field 
\begin{align}
\lambda_{\omega\Omega,33}&=\frac{F_\pi^2}4
S_\Theta^2 S_P^2+\frac{S_P^2}{e^2r^2}S_\Theta^2\Big(P_\theta^2+r^2P_r^2\cr
&+S_P^2(\Theta_\theta^2+r^2\Theta_r^2)\Big).
\end{align}
Finally, the additional parts of the moments of inertia arising from
the external magnetic field are expressed as 
\begin{align}
\Delta\lambda_{\omega\omega,12}&=\frac{q_eB_M}{4e^2}
\left(4+q_eB_Mr^2s_\theta^2\right)S_P^4S_\Theta^2,\\
\Delta \lambda_{\omega\Omega,12}&=\frac{q_eB_M}{4e^2}
\left(4+q_eB_Mr^2s_\theta^2\right)S_P^4S_\Theta^2
\Theta_\theta,\\
\Delta\lambda_{\Omega\Omega,12}&=\frac{q_eB_M}{4e^2}
\left(4+q_eB_Mr^2s_\theta^2\right)S_P^2S_\Theta^2
\cr
&\times(P_\theta^2+S_P^2\Theta_\theta^2).
\end{align}

\end{appendix}

\end{document}